\documentstyle[aps]{revtex}
\begin{document}
\title{Electron-electron interactions, coupled plasmon-phonon modes,
and mobility in n-type GaAs}
\author{B.A. Sanborn}
\address{Semiconductor Electronics Division \\
National Institute of Standards and Technology, Gaithersburg, MD\,\, 20899}
\maketitle
\begin{abstract}
This paper investigates the mobility of electrons scattering from the
coupled system of electrons and longitudinal optical (LO) phonons in n-type
GaAs.
The Boltzmann equation is solved exactly for low electric fields
by an iterative method, including
electron-electron and electron-LO phonon
scattering dynamically screened in the random-phase-approximation (RPA).
The LO phonon self-energy is treated in the plasmon-pole approximation.
Scattering from ionized impurities screened in static RPA is calculated
with phase-shift cross sections,
and scattering from RPA screened deformation potential and piezoelectric
acoustic phonons is included in the elastic approximation.
The results show that
dynamic screening and plasmon-phonon coupling
significantly modify inelastic scattering at low temperatures and densities.
The effect on mobility is
obscured by ionized impurity scattering in conventionally doped material,
but should be important
in modulation doped structures.
For uncompensated bulk n-type GaAs,
the RPA phase-shift model for
electron-impurity scattering gives lower drift mobilities than
the standard Thomas-Fermi or
Born calculations
which are high compared to experiment.
Electron-electron scattering lowers the mobility further,
giving improved agreement with experiment though discrepancies persist
at high donor
concentrations ($n>10^{18}{\rm cm}^{-3}$).
When impurities are ignored, inelastic scattering from the
coupled electron-phonon system is the strongest scattering mechanism at
77 K for moderate doping. This result differs from the standard model
neglecting mode coupling and electron-electron scattering which has
the acoustic
modes dominant in this regime.

\end{abstract}

\section{Introduction}

Good quantitative agreement between previous
calculations and experiment on low-field
transport coefficients of most high-purity polar semiconductors\cite{RODE}
suggests that the basic electron interactions are well understood in these
materials. This is not the case for doped polar semiconductors.
A long-standing discrepancy between theoretical and experimental mobility
values
for moderately to heavily doped
 n-type GaAs has been noted by many authors\cite{LOW,M&B,LANCE,KUECH,FENG,WAL,%
CHATT,STRING,YAN,POTH,RODE2,MOORE}. The fact that calculated mobilities are
consistently too high, especially for moderate to
heavy doping and low temperatures,
has motivated suggestions that compensation effects\cite{WAL},
correlated impurity distributions\cite{LANCE,YAN},
modifications of screening theory for a multi-ion system\cite{M&B},
or multiple scattering from ionized impurities\cite{MOORE}
should be included to obtain accurate results.

At the same time, it is known that
electron-electron interactions
can strongly influence electron transport in doped semiconductors,
by screening matrix elements and renormalizing phonon
energies, and through electron-electron scattering.
More accurate treatments of electron-electron interactions frequently improve
agreement between theory and experiment. For example, screening of ionized
impurity potentials in the static RPA rather than the
commonly used linearized Thomas-Fermi approximation (LTFA) has been
shown\cite{SANBORN} to reduce theoretical electron mobility
in n$^{+}$ silicon by more than a factor of two at 77 K.
Including dynamically screened
electron-electron scattering gives a significant reduction in
drift velocity of hot electrons injected into n-GaAs
devices\cite{LUG}.
Calculations including the coupling between plasmons and longitudinal optical
(LO) phonons
give shorter
electron inelastic lifetimes\cite{KIM,JAL} and
enhanced hot-electron energy relaxation\cite{DASSAR,LEI} in n-type GaAs.
Previous attempts to include electron-electron scattering in GaAs mobility
studies have been limited to static Thomas-Fermi screening
of single-particle scattering\cite{CHATT}.
So far, a treatment of inelastic scattering from
the nonequilibrium coupled mode system in combination with
an accurate model of ionized impurity scattering
has not been attempted.

In reference \cite{DERIV}, the inelastic collision term in the
electron Boltzmann equation was derived by making the Born approximation and
the
random-phase-approximation (RPA)
for scattering from the coupled
electron-LO phonon system in a doped polar semiconductor.
The result is the
sum of an electron-electron collision term and an electron-LO phonon
collision term that includes plasmon-phonon mode coupling.
Both interactions are dynamically screened by only the electronic part of the
total dielectric function for the electron-phonon system. The present paper
describes an iterative method for exactly solving the
low-field Boltzmann equation
including the dynamically screened
inelastic collision terms for arbitrary electron degeneracy  and spherical
energy surfaces.

For steady-state transport,
electron-plasmon scattering is correctly accounted for by dynamically screening
the electron-electron collision term\cite{DERIV}.
Electron-plasmon and
electron-electron scattering events
conserve total electron momentum and energy in direct gap doped semiconductors
since Umklapp processes are negligible in these materials.
Therefore, the purely electronic scattering events cannot by themselves degrade
an electrical curent,
if spherical energy surfaces are assumed, and they
 merely rearrange the nonequilibrium electron momentum distribution produced
by an applied field, thereby changing the probabilities for other scattering
processes.

In the present work, the effect of electronic scattering
is accounted for
by solving the Boltzmann equation for the
nonequilibrium electron distribution with electron-electron,
electron-phonon, and electron-impurity
scattering
treated simultaneously.
An isotropic effective electron mass and parabolic energy bands
are assumed.
The phonon distribution is approximated
 by its equilibrium form, and the plasmon-pole
approximation\cite{PP,PPSE} is used in the LO phonon self-energy.
 Numerical results are presented for
electron drift mobility in uncompensated n-type GaAs, including also electron
scattering from piezoelectric and
deformation potential acoustic phonons screened in the static
RPA. The calculations for conventionally doped material include phase-shift
scattering from ionized impurity potentials screened in the static RPA.
This method
gives lower mobilities than LTFA or Born calculations
of electron-impurity scattering, in better agreement with
experiment. Results for n-type GaAs with no impurities present show that
electron-electron scattering and mode coupling rather dramatically alter
inelastic scattering at 77 K, suggesting that these effects are significant
for modulation doped material at low temperature.

This paper is organized as follows. Section II discusses the screening
functions appropriate for the various electron scattering mechanisms included
in this study.
Section III describes the iterative method used to solve the Boltzmann
equation. Section IV presents the numerical results with discussion emphasizing
the differences obtained by using RPA vs LTFA screening, and by including
electron-electron scattering and plasmon-LO phonon coupling.

\section{Screening in a multicomponent system}

In reference \cite{DERIV}, it was shown that the nonequilibrium character
of screened electron interactions is essential to obtaining the correct
form of the collision terms in the Boltzmann equation, which contain
nonequilibrium distribution functions for the system's excitations. When
this is done, there remain dielectric functions which screen
the interaction matrix elements in the collision terms.
When the Boltzmann equation is linearized with respect to electric field
strength, only terms with equilibrium dielectric functions screening the
matrix elements in the collision terms
are nonvanishing. Therefore, consider the equilibrium screening process.
The interaction between two test charges in a solid can be expressed as the
bare Coulomb interaction $v_{q}=4\pi e^{2}/q^{2}$
screened by the total dynamic dielectric function
$\epsilon_{T}(q,\omega)$ for the system, including contributions from both the
conduction and lattice charges.
The frequency dependence of the screened
interaction is determined by the energy difference between the
scattering electron's
final and initial states.

In the RPA,
the electron-electron interaction is equivalent to the
test charge-test charge interaction.
In the same approximation, the total interaction $v_{q}/\epsilon_{T}
(q,\omega)$ is separable into a purely electron-electron interaction plus an
electron-phonon interaction including a phonon self-energy term, each of these
interactions being screened by only the electronic part of $\epsilon_{T}$.
The screened electron-phonon interaction $V_{sc-ph}$ includes the interactions
with {\em all} relevant phonons. It can be determined\cite{MAHAN} by
subtracting from the total
interaction the purely electronic interaction $v_{q}^{\infty}/
\epsilon(q,\omega)$,
\begin{eqnarray}
V_{sc-ph}(q,\omega)=\frac{v_{q}}{\epsilon_{T}}-\frac{v_{q}^{\infty}}{\epsilon}\\
=\frac{V_{ph}}{\epsilon[1-(v_{q}^{\infty}+V_{ph})P]}\,,
\end{eqnarray}
where $v_{q}^{\infty}=v_{q}/\epsilon_{\infty}$,
$\epsilon(q,\omega)=1-v_{q}^{\infty}P(q,\omega)$ is the electronic RPA
dielectric function, and $P$ denotes the polarization of the noninteracting
electron gas in this paper.
The bare electron-phonon interaction $V_{ph}$ is the sum over all phonon modes
$\lambda$ of the product of the phonon Green's function $D_{\lambda}$ and the
interaction matrix element $M_{\lambda}$ squared,
$V_{ph}=\sum_{\lambda}|M_{\lambda}|^{2}D_{\lambda}(q,\omega)$.
This method was used in reference \cite{DERIV} in order to determine
the screened electron-LO phonon interaction
$V_{sc-LO}$ for the case when only LO phonons are present ($V_{ph}=V_{LO}$).

When acoustic phonons are present as well,
$V_{ph}=V_{LO}+V_{ac}$ and
the total screened electron-phonon interaction $V_{sc-ph}$
can be separated into
$V_{sc-LO}$ (as determined before) plus the screened electron
interactions with acoustic phonons: $V_{sc-ph}=V_{sc-LO}+V_{sc-ac}$.
Solving for $V_{sc-ac}$,
\begin{eqnarray}
V_{sc-ac}(q,\omega)&=&\frac{V_{LO}+V_{ac}}{\epsilon[1-(v_{q}^{\infty}+V_{LO}+V_{ac})P]}-
\frac{V_{LO}}{\epsilon[1-(v_{q}^{\infty}+V_{LO})P]}\\
&=&\frac{V_{ac}}{(\epsilon-V_{LO}P-V_{ac}P)(\epsilon-V_{LO}P)}\,.
\end{eqnarray}
Since the acoustic phonon frequencies are small for relevant $q$ values,
$V_{sc-ac}$ is evaluated in the static limit. Using $V_{LO}(q,\omega=0)=v_{q}
(\epsilon_{0}^{-1}-\epsilon_{\infty}^{-1})$, which implies
that $\epsilon-V_{LO}P=
1-v_{q}P/\epsilon_{0}$, we have
\begin{equation}
V_{sc-ac}=\frac{\epsilon_{0}^{2}V_{ac}}{(\epsilon_{0}-v_{q}P)^{2}
[1-\epsilon_{0}V_{ac}P/(\epsilon_{0}-v_{q}P)]}\,.
\end{equation}
Thus, $\epsilon_{0}$ appears in the acoustic mode screening function. The
acoustic phonon self-energy $\epsilon_{0}V_{ac}P/(\epsilon_{0}-v_{q}P)$
is neglected
in the present work.

The electron-impurity interaction may also be viewed as a test charge-test
charge interaction.
Since the scattering
process is assumed to be nearly elastic, the relevant screening function is
$\epsilon_{T}(q,\omega)$ evaluated at $\omega=0$. In this case, $\epsilon_{T}$
should include the electronic and LO phonon parts plus a contribution from
the piezoelectric phonon interaction. The piezoelectric contribution is
assumed to be small and is neglected here so that the screening function for
the electron-impurity interaction is $\epsilon_{0}-v_{q}P$, the same as for
the acoustic modes.

\section{Numerical method}

In order to solve the electron Boltzmann equation
for a weak  electric field ${\bf F}$,
the nonequilibrium distribution function $f({\bf k}_{i})$
may be expanded around
the equilibrium solution
$f^{0}_{i}=\{1-\exp[\beta(E_{i}-\mu)]\}^{-1}$
to first order in ${\bf F}$,
\begin{equation}
f({\bf k}_{i}) = f^{0}_{i}+f^{0}_{i}(1-f^{0}_{i})x_{i}\psi_{i}\,,
\label{linear}
\end{equation}
where $x_{i}$ is the cosine of the angle between ${\bf F}$ and ${\bf k}_{i}$,
and $\psi_{i}=\psi(k_{i})$
is a function that is linear in $F$, but is otherwise unknown
 and to be determined by solving the Boltzmann equation.

Rode\cite{RODE} devised and implemented an iterative procedure
for solving the Boltzmann equation to determine
$f({\bf k})$ to first order in ${\bf F}$
for the case of spherical energy surfaces, including electron-phonon and
electron-impurity scattering only.
Assuming the linear form of $f$ (\ref{linear}) and using the law of cosines
in the electron Boltzmann equation (equation (40) of reference \cite{DERIV})
gives an integral equation for
the nonequilibrium electron distribution function\cite{RODE}.
\begin{eqnarray}
\frac{eF}{\hbar}\frac{\partial f^{0}_{k}}{\partial k}x_{k}
=-\{\stackrel{.}{f}_{k}\}^{linear}_{coll}&=&
\frac{x_{k}}{\Omega}\sum_{{\bf p}}
g_{p}x_{kp}\bigl\{W({\bf k,p})[1-f^{0}_{k}]+W({\bf k,p})f^{0}_{k}\bigr\}
\nonumber \\
&-&g_{k}\frac{x_{k}}{\Omega}\sum_{{\bf p}}\bigl\{W({\bf k,p})[1-f^{0}_{p}]+
W({\bf p,k})f^{0}_{p}\bigr\} \,,\label{linB}
\end{eqnarray}
where $g_{k}=f^{0}_{k}(1-f^{0}_{k})\psi_{k}$,
$x_{kp}$ is the cosine of the angle between ${\bf k}$ and ${\bf p}$,
and $W({\bf k,p})$ is the total
differential scattering rate for electronic transitions from state $|{\bf k}
\rangle$ to state $|{\bf p}\rangle$ due to scattering from ionized impurities
and phonons.
In the present work, $W({\bf k,p})$ for
the interaction with
LO phonons is modified to include the plasmon-phonon coupling as
described in reference\cite{DERIV}.
Approximating the phonon distribution function by its equilibrium form
$N^{0}(\omega)=(e^{\beta\hbar\omega}-1)^{-1}$, and using the plasmon-pole
approximation on the phonon self-energy gives
\begin{eqnarray}
W^{LO}({\bf k,p})=\frac{-2M_{q}^{2}}{\Omega\hbar|\epsilon(q,\omega_{k,p})|^{2}}
[N^{0}(\omega_{k,p})+1]{\rm Im}[D^{+}(q,\omega_{k,p})+D^{-}(q,\omega_{k,p})]\\
{\rm Im}[D^{\pm}(q,\omega)]=\mp\frac{\pi \omega_{TO}(\omega^{2}-\tilde{\omega}
_{p}^{2})}{\hbar \omega_{\pm}(\omega_{+}^{2}-\omega_{-}^{2})}[\delta(\omega+
\omega_{\pm})-\delta(\omega-\omega_{\pm})] \,,\label{IMDPM}
\end{eqnarray}
where $M_{q}^{2}=v_{q}(\epsilon_{\infty}^{-1}-\epsilon_{0}^{-1})\hbar\omega_
{LO}^{2}/2\omega_{TO}$ and
 the frequencies $\tilde{\omega}_{p}$, $\omega_{+}$, and $\omega_{-}$ are
given by
\begin{eqnarray}
\tilde{\omega}_{p}^{2}=\omega_{p}^{2}[1-\epsilon^{-1}(q,0)]^{-1}\\
\omega^{2}_{\pm}=\frac{1}{2}\left\{\omega^{2}_{LO}+\tilde{\omega}^{2}_{p}\pm
\left[(\omega_{LO}^{2}-\tilde{\omega}^{2}_{p})^{2}+4\omega_{p}^{2}(\omega_{LO}^
{2}-\omega_{TO}^{2})\right]^{1/2}\right\}\,.\label{OMEGAPM}
\end{eqnarray}
The weight factors $(\omega_{\pm}^{2}-\tilde{\omega}_{p}^{2})(\omega_{+}^{2}-
\omega_{-}^{2})^{-1}$ in Im$[D^{\pm}]$ give the phonon strength in each of
the hybrid $\omega_{\pm}$ modes, so that the differential scattering rate
$W^{LO}$ is the rate for scattering from only the phonon component of the
hybrid modes.
The screening function for the LO phonons is the electronic part of the
total dielectric function, $\epsilon(q,\omega)=1-v_{q}^{\infty}P(q,\omega)$.

Rode's iterative method can be generalized
to include electron-electron scattering by adding the linearized version of
the electron-electron collision term
$\{\stackrel{.}{f}\}^{ee}_{coll}$
(equation (47) of reference 1) to the right-hand side of (\ref{linB})
before solving for $\psi_{k}$.
The appendix shows how this is done by again using the law of cosines and
simplifying
$\{\stackrel{.}{f}\}^{ee}_{coll}$. Then,
the nonequilibrium
function $\psi$ satisfies the integral equation,
\begin{equation}
\psi(k)=\left[\nu _{inel}
 [\psi(k)] - \frac{e F}{\hbar}
\frac{\partial f^{0}_{k}}{\partial k}\right]
\left[ \frac{1}{\tau_{ee}(k)}+f^{0}_{k}(1-f^{0}_{k})\left(\frac{1}{\tau
_{el}(k)}
+\frac{1}{\tau
_{LO}(k)} \right)\right]^{-1}, \label{ITERATE}
\end{equation}
with the terms defined in this section.

For elastic scattering processes, the identities $k=p$ and $W({\bf k,p})=
W({\bf p,k})$ apply in equation (\ref{linB}) so that
the elastic scattering rate has the form
\begin{equation}
\tau^{-1}_{el}(k)=\frac{1}{\Omega}\sum_{{\bf p}}\left(1-x_{kp}\right)W_{el}
({\bf k},{\bf p}) \,.
\end{equation}
Interactions with acoustic phonons are approximated as elastic so that
$\tau^{-1}_{el}$ is composed of the rates for acoustic deformation potential,
piezoelectric, and ionized impurity scattering,
$\tau^{-1} _{el}(k)=\tau^{-1}_{dp}(k)+\tau^{-1}_{pe}(k)+\tau^{-1}_{ii}(k)$.
The acoustic mode scattering rates are calculated as in reference \cite{RODE}
except the interactions are screened in the static RPA as described in the
previous section.
\begin{eqnarray}
\frac{1}{\tau_{dp}(k)}=\frac{E_{1}^{2}k_{B}Tm^{*}}{4\pi\hbar^{3}c_{l}k^{3}}
\int_{0}^{2k}dq\frac{q^{3}}{(1-v_{q}P(q,0)/\epsilon_{0})^{2}} \label{DP} \\
\frac{1}{\tau_{pe}(k)}=\frac{e^{2}P_{pe}^{2}k_{B}Tm^{*}}{\epsilon_{0}\hbar^{3}
k^{3}}\int_{0}^{2k}dq\frac{q}{(1-v_{q}P(q,0)/\epsilon_{0})^{2}} \,,
\end{eqnarray}
where $E_{1}$ is the acoustic deformation potential, $c_{l}$ is the spherically
averaged elastic constant for longitudinal modes, and $P_{pe}$ is the
piezoelectric
coefficient.

The electron-impurity scattering rate $1/\tau_{ii}^{ps}$
is calculated with the phase shifts
$\delta_{l}(k)$ determined by
numerically solving the radial Schr\"{o}dinger equation
with an impurity potential screened with the static RPA total dielectric
function $\epsilon_{T}(q,0)$\cite{SANBORN}. The results
presented below include a comparison with the
Born approximation $1/\tau_{ii}^{Born}$.
\begin{eqnarray}
\frac{1}{\tau_{ii}^{ps}(k)}=\frac{4\pi\hbar}{km^{*}}N_{i}\sum_{l=0}^{\infty}
(l+1)sin^{2}[\delta_{l}(k)-\delta_{l+1}(k)]\label{TIMPPS}\\
\frac{1}{\tau_{ii}^{Born}(k)}=\frac{4\pi^{4}m^{*}}{\epsilon_{0}^{2}(\hbar k)
^{3}}N_{i}\int_{0}^{2k}dq\frac{1}{q}\frac{1}{(1-v_{q}P(q,0)/\epsilon_{0})^{2}}
\label{TIMPBORN} \,,
\end{eqnarray}
where $N_{i}$ is the ionized impurity concentration.

The LTFA is made by using the temperature-dependent Thomas-Fermi dielectric
function $\epsilon_{LTFA}=1+q_{TF}^{2}/q^{2}$, where $q_{TF}^{2}=4\pi e^{2}/
\epsilon(\partial n/\partial \mu)$ and $\mu$ is the chemical
potential. The dielectric constant $\epsilon$ is $\epsilon_{\infty}$ for
electron-electron or electron-LO phonon scattering, while $\epsilon=\epsilon_
{0}$ for the screened impurity and acoustic phonon potentials.
 Using the LTFA in equation (\ref{TIMPBORN})
gives the familiar Brooks-Herring formula.

The inverse lifetimes $\tau^{-1}_{ee}(k)$ and $\tau^{-1}_{LO}(k)$ describe
the rate at which the nonequilibrium population of state $|{\bf k}\rangle$
decays due to inelastic scattering with the equilibrium population of the
other states.
For electron-electron collisions,
\begin{equation}
\tau^{-1}_{ee}(k)=\frac{8e^{4}f_{k}}{\epsilon_{\infty}^{2}\hbar k}\int_{0}
^{\infty}dp\,p(1-f_{p})[N^{0}(\omega_{k,p})+1]\int_{\mid k-p \mid}^{k+p}
\frac{dq}{q^{3}}\,\frac{{\rm Im}[P (q,\omega_{k,p})]}{|\epsilon(q,\omega_
{k,p})|^{2}} \,.
\end{equation}
For scattering with the LO phonon component of the hybrid modes,
\begin{eqnarray}
\tau^{-1}_{LO}(k)=
\sum_{\lambda=\pm}
\lambda&\Biggl\{\Biggr.&\int_{|k-k_{\lambda}^{-}|}^{k+k_{\lambda}^{-}}
\,dq I(q,\omega_{\lambda})
\Bigl[N^{0}(\omega_{\lambda})+1-f^{0}_{k_{\lambda}^{-}}\Bigr]
\Theta(E_{k}-\hbar\omega_
{\lambda}) \nonumber \\
& &+\int_{|k-k_{\lambda}^{+}|}^{k+k_{\lambda}^{+}}
\,dq I(q,\omega_{\lambda})
\Bigl[N^{0}(\omega_{\lambda})+f^{0}_{k_{\lambda}^{+}}\Bigr]\Biggr\} \,,
\label{TAULO}
\end{eqnarray}
where
\begin{eqnarray*}
k_{\lambda}^{\pm}&=&\sqrt{k^{2}\pm2m^{*}\omega_{\lambda}/\hbar} \,, \\
I(q,\omega_{\lambda})&=&\frac{e^{2}\omega_{LO}m^{*}}{q\omega_{\lambda}\hbar^{2}k
}
(\frac{1}{\epsilon_{\infty}}-\frac{1}{\epsilon_{0}})\mid\epsilon(q,\omega
_{\lambda})\mid^{-2}
\frac{(\omega^{2}_{\lambda}-\tilde{\omega}_{p}^{2})}{(\omega_{+}^{2}-
\omega_{-}^{2})} \,.
\end{eqnarray*}

Since the mobility is measured on a system with {\em all} of the electrons
out of equilibrium, the linear integral equation (\ref{ITERATE}) also
contains the functional $\nu_{inel}[\psi]$ describing
the rate of change of the
equilibrium occupation of the state $|{\bf k}\rangle$ due to inelastic
scattering with the nonequilibrium population of the other states.
This term is composed of electron-electron and electron-LO
phonon parts: $\nu_{inel}=\nu_{ee}+\nu_{LO}$.
The contribution from electron-electron scattering is
\begin{eqnarray}
&\nu _{ee}&[\psi]=
\frac{-4e^{4}f^{0}_{k}}{\epsilon_{\infty}^{2}\hbar k^{2}}
\int_{0}^{\infty}dp\,(1-f^{0}_{p})
\int_{\mid k-p \mid}
^{k+p}\frac{dq}{q^{3}}\mid \epsilon(q,\omega_{k,p})\mid^{-2} \nonumber \\
&\times&\Biggl\{\Biggr.\psi_{p}(k^{2}+p^{2}-q^{2})[N^{0}
(\omega_{k,p})+1]{\rm Im}[P(q,\omega_{k,p})]  \nonumber \\
&+&\frac{m^{*}p}{2\pi\hbar^{2}q^{2}}(k^{2}-p^{2}+q^{2})
\Bigl[\Bigr.z^{-}\int_{|z^{-}|}
^{\infty}ds\,\psi_{s}f^{0}_{s}(1-f^{0}_{s^{+}})
-z^{+}\int_{|z^{+}|}^{\infty}ds\,\psi_{s}f^{0}_{s^{-}}(1-f^{0}_{s})
\Bigl.\Bigr]
\Biggl.\Bigg\}\,,
\end{eqnarray}
where $z^{\pm}=(k^{2}-p^{2}\pm q^{2})/2q$ and
$s^{\pm}=\sqrt{s^{2}\pm(k^{2}-p^{2})}$.
The contribution from the LO phonon strength in the hybrid modes is
\begin{eqnarray}
\nu_{LO}[\psi]=
\sum_{\lambda=\pm}
\lambda&\Biggl\{\Biggr.&\int_{|k-k_{\lambda}^{-}|}^{k+k_{\lambda}^{-}}
\,dq I(q,\omega_{\lambda})
g(k_{\lambda}^{-})\frac{(k^{2}+(k_{\lambda}^{-})^{2}-q^{2})}{2kk_{\lambda}
^{-}}
\Bigl[N^{0}(\omega_{\lambda})+f^{0}_{k}\Bigr]
\Theta(E_{k}-\hbar\omega_
{\lambda}) \nonumber \\
& &+\int_{|k-k_{\lambda}^{+}|}^{k+k_{\lambda}^{+}}
\,dq I(q,\omega_{\lambda})
g(k_{\lambda}^{+})\frac{(k^{2}+(k_{\lambda}^{+})^{2}-q^{2})}{2kk_{\lambda}^{+}}
\Bigl[N^{0}(\omega_{\lambda})+1-f^{0}_{k}\Bigr]\Biggr\}\,.\label{NULO}
\end{eqnarray}

Plasmon-phonon coupling may be neglected by setting $\omega_{+}$ equal to
$\omega_{LO}$ and $\omega_{-}$ equal to $\tilde{\omega}_{p}$ in equations
(\ref{TAULO}) and (\ref{NULO}).

\section{Results and discussion}
The numerical results for the drift mobility were obtained by solving
the iterative
equation (\ref{ITERATE})
with (\ref{linear}) for the
linearized nonequilibrium electron distribution in order
to determine the average velocity per unit electric field.
The effective electron mass value $m^{*}=0.07m_{e}$ was used throughout
the calculations.
The other material parameters used were taken from reference
\cite{NAG}.
Results for uncompensated n-type GaAs are presented for conventionally doped
material (figures 1 through 4)
 and for a doped system with no impurities present, an idealized model
of modulation doped material (figures 5 through 7).

\subsection{RPA vs LTFA screening}
The LTFA is an approximation to
the static RPA
dielectric function which does not account for the polarization
of the screening electrons by the colliding electron, but is equivalent in
the limit of small momentum transfer, $q$.
Figures 1 and 2
show the carrier concentration dependence of electron mobility in
uncompensated n-type GaAs
at 300  and 77 K, calculated with statically screened electron-phonon and
electron-impurity interactions.
Scattering from LO phonons as well as the acoustic
piezoelectric and deformation potential modes
was included, but
electron-electron scattering and plasmon-phonon coupling were neglected.
Each figure shows calculations made by using either
the phase-shift (\ref{TIMPPS})
or Born (\ref{TIMPBORN}) electron-impurity scattering rate, and either
the temperature-dependent static RPA or LTFA dielectric function  in
equations (\ref{DP}-\ref{TIMPBORN}).
The standard LTFA Born (Brooks-Herring) calculation with unscreened phonons
is also shown.
Compared to the case of n-type Si\cite{SANBORN}, a relatively small correction
to mobility in n-type GaAs is obtained by using RPA screening rather than
LTFA. Except for the unscreened phonon case at 300 K, the combination of RPA
screening of impurities and phonons with the phase-shift
electon-impurity scattering rate gives
the lowest mobilities, while the LTFA Born calculations give values that
are higher by at most $16\%$ at 300 K and $41 \%$ at 77 K. A
cancellation of errors made by using the LTFA Born electron-impurity
interaction
and unscreened electron-phonon interactions at 300 K makes this result
similar in magnitude to the RPA Born curve.
Dynamic RPA screening of LO phonons changes the mobility very little from the
static RPA case when electron-impurity scattering is included.

\subsection{Electron-electron scattering}
The effect of electron-electron scattering on mobility
in doped semiconductors is limited by three factors.
1) The screened interaction between two electrons
is considerably weaker than the Coulomb interaction. 2) When Umklapp
electronic scattering processes are negligible and energy surfaces are close
to spherical, electron-electon scattering conserves total electron
momentum and produces only an indirect effect on other scattering processes
by redistributing the electron momenta. 3) The Pauli principle restricts the
fraction of electrons that can participate in energy-conserving
 electron-electron
scattering processes to a number that vanishes in the degenerate limit.
Nevertheless,
electron-electron scattering was shown\cite{APPEL} to have
an important effect on transport in doped semiconductors away from the
degenerate limit.
Chattopadhyay\cite{CHATT} studied its effect on mobility in n-type GaAs
by using
a variational solution of the Boltzmann equation including electron-electron
scattering statically screened in LTFA with $\epsilon_{0}$. He found that,
for $n=10^{16}$ cm$^{-3}$, electron-electron scattering reduces the mobility
by about $10\%$ at 80 K. The reduction increases to $20 \%$ when
$\epsilon_{\infty}$ rather than $\epsilon_{0}$ is used in the LTFA screened
electron-electron interaction, as discussed in reference \cite{DERIV}.
Figures 3 and 4 show the drift mobility calculated by iteratively solving
the Boltzmann equation at 300 and 77 K, including electron-electron
scattering in addition to the electron-impurity and electron-phonon
mechanisms already discussed. The Pauli principle restriction is clearly
seen from the vanishing of the electron-electron contribution at higher
concentrations. Also shown in figures 3 and 4 are experimental
data\cite{FENG,STRING,HOUNG} for the Hall mobility of n-type
GaAs. No attempt has been made in the calculations to include the Hall factor
relating the drift to Hall mobilities.
The Hall factor is generally larger than unity so that the Hall mobility is
larger than the drift mobility, but its magnitude depends on doping density
and temperature, approaching unity in the degenerate limit.
 Despite the reduced mobility values
compared to previous calculations (see e.g. \cite{RODE,LOW,LANCE,WAL}),
 a disagreement between theory and
experiment persists for carrier concentrations above $n=10^{18}$ cm$^{-3}$
at both temperatures considered.

\subsection{Coupled plasmon-phonon modes}
The uncoupled plasmon temperature $\hbar \omega _{p}/k_{B}$ varies as
$n^{1/2}$;
for n-type GaAs, it
ranges from 50 K to 1594 K for electron concentrations of $10^{16}$ cm$^{-3}$
to
$10^{19}$ cm$^{-3}$.
The plasmon temperature
crosses the LO phonon temperature of 419 K at $\sim 7 \times
10^{17}$ cm$^{-3}$. In this concentration region,
the electronic and lattice
excitations hybridize to form the normal modes of mixed electron-phonon
character.
The other energy relevant to electron mobility is the Fermi energy $E_{F}$,
which varies as $n^{2/3}$. For n-type GaAs, $E_{F}/k_{B}$ ranges from 29 K to
2930 K for $n=10^{16}$ cm$^{-3}$ to $10^{19}$ cm$^{-3}$ and crosses the
LO phonon temperature at $\sim 6 \times 10^{17}$ cm$^{-3}$.

In the plasmon-pole model, the phonon spectral function $-\pi ^{-1}
{\rm Im}[D(q,\omega)]$ has two delta-function peaks (see equation
(\ref{IMDPM}))
with frequencies $\omega_{+}$ and $\omega_{-}$ given by
(\ref{OMEGAPM}).
As discussed in reference \cite{DERIV}, mobilities
at low densities and temperatures are expected to be reduced when
plasmon-LO phonon coupling is included, because of
increased low-energy electron-phonon scattering due to the
$\omega_{-}$ mode.
Figure 5 shows that this is in fact the case.
The mobilities in this figure were calculated by neglecting all scattering
mechanisms except electron-LO phonon scattering.
For a wide range of densities at 77 K,
the LO phonon limited mobility
including mode coupling and dynamic RPA screening (solid curve)
is significantly lower than the mobilites calculated by neglecting these
effects. The dip observed below $n=10^{18}$ cm$^{-3}$ in all the figure 5
mobility curves is related to $E_{F}$ crossing the LO phonon emission
threshold.
An interesting though minor effect appears in figure 3,
where the mobility including mode coupling is actually higher at low doping
levels and 300 K than the mobility neglecting the coupling, when
electron-impurity is included.

Since mobility is limited primarily by
scattering from ionized impurities at low temperatures in conventionally
doped GaAs,
the mode-coupling effect is obscured in this case.
Nevertheless, it could be important for mobility in
modulation doped structures where  the effects of impurity scattering are
greatly reduced because of the large separation between dopants and carriers.
Figure 6 plots the
electron drift mobility as a function of concentration at 77 K
when electron-impurity scattering is neglected. The figure shows that
scattering from the acoustic (piezoelectric and deformation potential) modes
is stronger than scattering from LO phonons
when electron-electron scattering is neglected, even if mode coupling is
included.
However, in the moderate doping regime, inelastic scattering is stronger than
scattering from acoustic phonons when both
electron-electron scattering and mode coupling are included.
This result is notable especially because the value used in the calculation
for the acoustic deformational potential $E_{1}=12$ eV
is higher than some others which appear in the literature.
Figure 7 plots the temperature dependence of the mobility from 50 to 150 K
for $n=10^{17}{\rm cm}^{-3}$ when electron-impurity
scattering is neglected. The figure shows that electron-electron scattering
and mode coupling effects should be significant even at the lower temperatures
in situations where impurity effects are minimal.

\section{Conclusions}
The mobility calculations presented in this paper show that a more accurate
treatment of electron interactions in n-type GaAs yields results that differ
from simple models, especially at moderate doping levels in
conventionally doped material and for low densities and temperatures
in modulation-doped material.
For conventional n-type GaAs,
RPA screening of phonons and impurities combined with phase-shift
electron-impurity scattering rates give lower drift mobilities
 than LTFA screening or Born calculations.
A convergent iterative method has been developed for solving the Boltzmann
equation including electron-electron scattering.
Including this scattering mechanism
further reduces
the mobility except at high doping levels, so that the full calculation gives
better agreement with experimental Hall mobilities than previous results.
In order to say definitively that the long-standing discrepancy
between theory and experiment is resolved
in the middle concentration  regime, the Hall mobility
should be calculated with the present model.
At the highest concentrations, the theoretical mobility for uncompensated
n-type GaAs is larger than experimental values.

The effects of dynamic screening, mode coupling, and electron-electron
scattering are
not very important for mobility in conventionally
doped GaAs where scattering from ionized impurities dominates. However,
the calculations neglecting impurities suggest that these effects
could significantly affect mobilities in modulation-doped structures.
Comparison with experiment would require a more realistic treatment including
the effects of confinement on phonon spectra, the two-dimensional nature
of the electron gas in these structures,
as well as the effects of impurities, alloy disorder, and interface roughness.
Also, the effect of Landau damping
of the hybrid modes, which is neglected in the plasmon-pole model
for the phonon self-energy,
should be investigated.
 Good candidates for study would be Al$_{x}$Ga$_{1-x}$As,
since the Fr\"{o}hlich coupling increases with $x$\cite{ADACHI}, or the more
strongly polar materials CdTe, ZnSe, or ZnTe.
Finally, the fact that inelastic scattering is stronger
compared to acoustic phonon scattering than the standard model
predicts  for doped material at low temperature suggests that
the inelastic component could influence determinations of the deformation
potential constant from mobility studies, and motivates a reevaluation of
$E_{1}$ in n-type GaAs at low temperatures.

\acknowledgements
I thank J.R. Lowney, G.W. Bryant, and S. Das Sarma for helpful conversations.

\newpage

\appendix
\section*{Iterative method for electron-electron scattering}
Rode's method\cite{RODE} is generalized by inserting the linearized form of
$f$ (\ref{linear}) into the electron-electron collision term
(equation (47) of reference \cite{DERIV}) and
using the identity
\begin{equation}
\delta(E_{1}+E_{2}-E_{3}-E_{4})\,
\left[f_{1}^{0}f_{2}^{0}(1-f_{3}^{0})(1-f_{4}^{0})-(1-f_{1}^{0})
(1-f_{2}^{0})f_{3}^{0}f_{4}^{0}\right]=0,
\end{equation}
giving the linearized collision term,
\begin{eqnarray*}
\{\stackrel{.}{f}({\bf k}_{1})\} ^{ee}_{coll}
&=&\frac{1}{\Omega^{3}}
\sum _{2,3,4} |M_{3,1}|^{2}  \delta(E_{1}+E_{2}-E_{3}-E_{4})
\delta ({\bf k}_{1}+{\bf k}_{2}-{\bf k}_{3}-{\bf k}_{4}) \nonumber \\
&\times
&f^{0}_{1}f^{0}_{2}(1-f^{0}_{3})(1-f^{0}_{4})[x_{1}\psi_{1}+x_{2}\psi_{2}
-x_{3}\psi_{3}-x_{4}\psi_{4}]\,.
\end{eqnarray*}
This symmetric form allows a simple application of the law of cosines to
eliminate $x_{i}$ for $i=2$ to $4$ in favor of $x_{1i}$, the cosine of the
angle between ${\bf k}_{1}$ and ${\bf k}_{i}$.
\begin{eqnarray}
\{\stackrel{.}{f}({\bf k}_{1})\} ^{ee}_{coll}
&=&\frac{x_{1}}{\Omega^{3}}
\sum _{2,3,4} |M_{3,1}|^{2}  \delta(E_{1}+E_{2}-E_{3}-E_{4})
\delta ({\bf k}_{1}+{\bf k}_{2}-{\bf k}_{3}-{\bf k}_{4}) \nonumber \\
&\times &f^{0}_{1}f^{0}_{2}(1-f^{0}_{3})(1-f^{0}_{4})[\psi_{1}+
x_{12}\psi_{2}
-x_{13}\psi_{3}-x_{14}\psi_{4}]\,.
\end{eqnarray}
Now use the delta function for momentum conservation to sum over ${\bf k}_{4}$
and introduce the variable ${\bf q}={\bf k}_{1}-{\bf k}_{3}={\bf k}_{4}-{\bf k}
_{2}$.
The ${\bf k}_{2}$ integration of
the $\psi_{1}$ and $\psi_{3}$ terms can be done analytically.
Making the
variable changes ${\bf k}_{1} \rightarrow {\bf k}, {\bf k}_{2} \rightarrow
{\bf s}, {\bf k}_{3}\rightarrow{\bf p}={\bf k-q}$, and ${\bf k}_{4}
\rightarrow{\bf s}+{\bf q}$, we have
\begin{eqnarray}
\{\stackrel{.}{f}({\bf k})\} ^{ee}_{coll}
=\frac{-x_{k}}{2\pi\Omega}\sum_{{\bf p}}\mid M_{k,p}\mid^{2}
(&\psi_{k}&-x_{kp}\psi_{p})
[N^{0}(\omega_{k,p})+1]{\rm Im}[P
(q,\omega_{k,p})]
f^{0}_{k}(1-f^{0}_{p}) \nonumber \\
+\frac{1}{\Omega^{2}}\sum_{{\bf s},{\bf p}} \mid M_{k,p}\mid^{2}
x_{ks}\,\psi_{s}
\Biggl\{\Biggr.&\delta&(E_{k}-E_{p}+\hbar\omega_{s,s+q})f^{0}_{k}f^{0}_{s}(1-f^{0}_{p})
(1-f^{0}_{s+q}) \nonumber\\
-&\delta&(E_{k}-E_{p}-\hbar\omega_{s,s-q})f^{0}_{k}f^{0}_{s-q}(1-f^{0}_{p})
(1-f^{0}_{s})\Biggl.\Biggr\}\,,
\end{eqnarray}
where ${\bf s}$ was changed to ${\bf s-q}$ in the last term, and
\begin{equation}
{\rm Im}[P(q,\omega)]=\frac{m^{*}}{2\pi\hbar^{4}\beta q}
\ln \left| \frac{1+\exp[-\beta(E^{+}-\mu)]}{1+\exp[-\beta(E^{-}-\mu)}
\right|\,,
\end{equation}
with $E^{\pm}=(E_{q}\pm\hbar\omega)^{2}/4E_{q}$.

The symmetry of the problem is best used by choosing bipolar
coordinates\cite{COMB} with the $z$ axis along ${\bf q}$ and the $xz$ plane
containing ${\bf k}$.
 The ${\bf p}$ integration variables can be chosen to include the momentum
transfer $q$,
\begin{displaymath}
\int d^{3}p
=2\pi \int_{0}^{\infty}dp\,p\int_{\mid k-p \mid}^{k+p}dq\,\frac{q}{k}\,,
\end{displaymath}
the factor of $2\pi$ coming from integration of the azimuthal angle around
${\bf k}$. The ${\bf s}$ integration has the form
$\int d^{3}s=\int d\phi \int dx_{sq} \int ds\,s^{2}$
where $\phi$ is the angle between the $({\bf k},-{\bf p})$ and $(-{\bf s},
{\bf k}_{4})$ plane. The $\phi$ integral
is simply done by again applying the law of
cosines,
\begin{displaymath}
\int_{0}^{2\pi}d\phi\,x_{ks}=2\pi \,x_{sq}\, x_{kq}\,.
\end{displaymath}
Then, using the energy delta functions to do the $x_{sq}$ integrals,
the linearized electron-electron collision integral reduces to
\begin{eqnarray}
\{\stackrel{.}{f}({\bf k})\} ^{ee}_{coll}&=&
\frac{-x_{k}f^{0}_{k}}{k(2\pi)^{3}}\int_{0}^{\infty}dp\,p(1-f^{0}_{p})
\int_{\mid k-p \mid}
^{k+p}dq\mid M_{k,p}\mid^{2} \nonumber \\
\times\Biggl\{\Biggl.&q&(\psi_{k}-x_{kp}\psi_{p})[N^{0}
(\omega_{k,p})+1]{\rm Im}[P(q,\omega_{k,p})]  \nonumber \\
&+&\frac{x_{kq}\,m^{*}}{2\pi\hbar^{2}}
\Bigl[\Bigr.z^{-}\int_{|z^{-}|}
^{\infty}ds\,\psi_{s}f^{0}_{s}(1-f^{0}_{s^{+}})
-z^{+}\int_{|z^{+}|}^{\infty}ds\,\psi_{s}f^{0}_{s^{-}}(1-f^{0}_{s})
\Bigl.\Bigr]
\Biggl.\Bigg\}\,, \label{finalee}
\end{eqnarray}
where $z^{\pm}=(k^{2}-p^{2}\pm q^{2})/2q$ and
$s^{\pm}=\sqrt{s^{2}\pm(k^{2}-p^{2})}$.

Adding (\ref{finalee}) to the right-hand side of (\ref{linB}) and solving for
$\psi_{k}$ gives equation (\ref{ITERATE}).

\begin{figure}
\caption{Comparison of RPA vs LTFA screening and phase-shift vs Born
electron-impurity cross sections at 300 K for n-type GaAs.
Static screening of LO, piezoelectric, and acoustic deformation potential
phonons as well as ionized impurities is included.
 The RPA phase-shift drift
mobility (solid curve) is shown with mobilities in the following
approximations:
RPA Born (dotted curve), LTFA phase-shift (long-dashed curve),
LTFA Born (short-dashed curve), and LTFA Born with unscreened phonons
(chain-dashed curve). Electron-electron scattering and plasmon-phonon mode
coupling has been neglected.}
\end{figure}

\begin{figure}
\caption{Comparison of RPA vs LTFA screening and phase-shift vs Born
electron-impurity cross
sections at 77 K for n-type GaAs. Calculated mobilities are represented as in
figure 1.}
\end{figure}

\begin{figure}
\caption{Effect of electron-electron scattering on electron
drift mobility in n-type
GaAs at 300 K. The dotted and dashed curves are the RPA phase-shift
calculation neglecting and including, respectively,
statically screened electron-electron scattering. The
solid curve includes dynamically screened electron-electron and electron-LO
phonon scattering with mode coupling.
Experimental Hall mobilities are indicated by  $\bigcirc$ (ref. [9])
and $\Box$ (ref. [27]).}
\end{figure}

\begin{figure}
\caption{Effect of electron-electron scattering on electron
drift mobility in n-type
GaAs at 77 K. Calculated mobilities are represented as in figure 4.
Experimental Hall mobilities are indicated by  $\bigcirc$ (ref. [9])
and $\triangle$ (ref. [12]).}
\end{figure}

\begin{figure}
\caption{Effects of screening and mode coupling
on LO phonon limited electron mobility in n-type GaAs
at 77 K.
The dynamically screened coupled mode
mobility calculation (solid curve) is compared to
mobilities determined by scattering from uncoupled
LO phonons screened with
LTFA(long-dashed curve), static
RPA (short-dashed curve), dynamic RPA (dotted curve), and unscreened
(chain-dashed curve). All other scattering mechanisms are neglected.}
\end{figure}

\begin{figure}
\caption{Comparison of mobilities limited by inelastic and acoustic phonon
scattering in n-type GaAs at 77 K.
The mobility including dynamically screened electron-electron
and electron-LO phonon scattering
including mode coupling (chain-dashed curve) is compared to
the same calculation except neglecting electron-electron scattering (dashed
curve), and to the mobility including only scattering from acoustic
(piezoelectric and deformation potential) phonons
screened with static RPA (dotted curve). The solid curve shows the
calculated mobility including all of the mechanisms. Electron-impurity
scattering is not included.}
\end{figure}

\begin{figure}
\caption{Temperature dependence of mobility in n-type GaAs for $n=10^{17}{\rm
cm}
^{-3}$. The mobility including dynamically screened electron-electron and
electron-LO phonon scattering including mode coupling (solid curve) is
compared to the results with static RPA screening of
electron-electron scattering (dotted curve), electron-electron scattering
neglected (dashed curve), and static RPA screening of LO phonons without
mode coupling (chain-dashed curve). Electron scattering from acoustic phonons
screened in static RPA is included and electron-impurity scattering is
neglected in all cases.}
\end{figure}

\end{document}